\documentclass[10pt,letterpaper]{article}
\usepackage{geometry}

\usepackage[utf8x]{inputenc}
\usepackage[english]{babel}
\usepackage[encapsulated]{CJK}

\usepackage{textcomp,marvosym}

\usepackage{fixltx2e}

\usepackage{amsmath,amssymb,bm}

\usepackage{cite}

\usepackage{nameref,hyperref}

\usepackage{microtype}
\DisableLigatures[f]{encoding = *, family = * }

\usepackage{color}
\usepackage[inline]{enumitem}

\usepackage[aboveskip=1pt,labelfont=bf,labelsep=period,justification=raggedright,singlelinecheck=off]{caption}

\makeatletter
\renewcommand{\@biblabel}[1]{\quad#1.}
\makeatother

\date{}

\usepackage{lastpage,fancyhdr,graphicx}
\usepackage{epstopdf}

\begin{document}

\begin{flushleft}
{\LARGE
\textbf\newline{Advances in using Internet searches to track dengue} 
}
\newline
\\
Shihao Yang\textsuperscript{1},
S. C. Kou \textsuperscript{1,*},
Fred Lu \textsuperscript{2},
John S. Brownstein\textsuperscript{2,3}, Nicholas Brooke \textsuperscript{4}
Mauricio Santillana\textsuperscript{2,3,
*}
\\
\bigskip
\textbf{1} Department of Statistics, Harvard University, Cambridge, MA, USA \\
\textbf{2} Computational Health Informatics Program, Boston Children's Hospital, Boston, MA, USA
\\
\textbf{3} Harvard Medical School, Boston, MA, USA\\
\textbf{4} The Synergist, Brussels, Belgium\\
\bigskip

%
%





* Corresponding authors: msantill@g.harvard.edu; kou@stat.harvard.edu

\end{flushleft}
\section*{Abstract}
Dengue is a mosquito-borne disease that threatens more than half of the world's population. Despite being endemic to over 100 countries, government-led efforts and mechanisms to timely identify and track the emergence of new infections are still lacking in many affected areas. Multiple methodologies that leverage the use of Internet-based data sources have been proposed as a way to complement dengue surveillance efforts. Among these, the trends in dengue-related Google searches have been shown to correlate with dengue activity. We extend a methodological framework, initially proposed and validated for flu surveillance, to produce near real-time estimates of dengue cases in five countries/regions: Mexico, Brazil, Thailand, Singapore and Taiwan. Our result shows that our modeling framework can be used to improve the tracking of dengue activity in multiple locations around the world.

\section*{Author Summary}
As communicable diseases spread in our societies, people frequently turn to the Internet to search for medical information. In recent years, multiple research teams have investigated how to utilize Internet users' search activity to track infectious diseases around our planet. In this article, we show that a methodology, originally developed to track flu in the US, can be extended to improve dengue surveillance in multiple countries/regions where dengue has been observed in the last several years. Our result suggests that our methodology performs best in dengue-endemic areas with high number of yearly cases and with sustained seasonal incidence.

\section*{Introduction}

Dengue fever poses a growing health and economic problem throughout the tropical and sub-tropical world. Dengue is today one of the fastest-growing and most important mosquito-borne viral diseases in the world, with an estimated 390 million dengue infections each year and threatening an estimated 3.9 billion people in 128 countries \cite{WHO2016dengue}. Infection often causes high fever and joint pain, and severe cases can lead to hemorrhage, shock and death. Dengue epidemics strain health services and lead to massive economic losses. 

Dengue mortality and morbidity both need to be addressed to reduce this heavy burden. The World Health Organization has called for better early case detection among other tactics to reduce dengue mortality, and to reduce dengue morbidity through improved outbreak prediction and detection through coordinated epidemiological and entomological surveillance.

However, a comprehensive, effective and reliable disease surveillance system for dengue is not yet available. Governments traditionally rely on hospital-based reporting, a method that is often lagged and limited with frequent post-hoc revisions, due to communication inefficiencies across local and national agencies and the time needed to aggregate information from the clinical to the state level \cite{runge2008does,madoff2011new}. This lack of timely information limits the identification and optimization of effective interventions. Measurement difficulties are compounded by the fact that a majority of dengue cases are asymptomatic \cite{world2009dengue}.

In this context, building an effective disease surveillance tool is essential to being able to identify and assess the severity of dengue outbreaks and to enable better assessment of the effectiveness of ongoing interventions. Such tool(s) should provide accurate and consistent measures of regional or national infection levels, be updated in near real-time, and be immune to bureaucratic or resource-related delays. To improve accuracy, these tools should use and link together multiple sources of information, using both traditional and non-traditional sources. 

Over the years, a broad range of traditional epidemiological methods have been proposed by research teams to fill this time gap of information by supplementing official case counts with now-cast predictions using historical dengue incidence data from previous seasons. Autoregressive models, such as Seasonal Autoregressive Integrated Moving Average (SARIMA) model, that take into account historical and seasonal patterns, have been shown to produce useful disease estimates, some including additional variables such as concurrent weather information \cite{johansson2016evaluating,promprou2006forecasting,luz2008time,choudhury2008forecasting,eastin2014intra}. Other studies have further examined climate-driven models, finding associations of seasonal and long-term weather patterns such as El Ni\~no with dengue levels in various countries \cite{johansson2009multiyear,hii2012forecast,hurtado2007short,lu2009time}.
In addition, various mechanistic models on the dynamics of dengue transmission have also been explored, with some recent promise \cite{aguiar2011role}. A comprehensive survey of these methods are given in Andraud et al. \cite{andraud2012dynamic}

In parallel and complementary to the aforementioned methodologies, the global spread of the Internet has opened up the opportunity to investigate whether users' activity patterns on Internet search-engines and social media platforms may lead to reasonable estimates of dengue infection levels \cite{chan2011using, gomide2011dengue, althouse2011prediction}. In theory, Internet search tracking is consistent, efficient, and reflects real-time population trends, giving it strong potential to supplement current epidemiological methods \cite{milinovich2014internet,madoff2011new}. Studies have previously demonstrated the feasibility of using Internet search data to track dengue case counts \cite{chan2011using,althouse2011prediction}. Google Dengue Trends (GDT), launched in 2011, was one of the first tools to quantitatively track dengue activity in multiple regions throughout the world by leveraging the aggregate Google search patterns of millions of users \cite{chan2011using}. Since its start, the methodology behind GDT has been updated to address flaws found in its sister effort, Google Flu Trends \cite{ginsberg2009detecting, cook2011assessing, olson2013reassessing, copeland2013google, lazer2014parable, santillana2014can, yang2015accurate, pollett2016evaluating, santillana2016perspectives}, before finally being discontinued in August 2015. An assessment of GDT in Mexico showed mixed prediction accuracy compared to official case counts, with strong correlation in some states \cite{gluskin2014evaluation}. 

Despite progress in the use of both dengue historical information (time series approaches \cite{johansson2016evaluating}) and real-time Internet searches for dengue tracking \cite{chan2011using}, an approach for accurate tracking of dengue by combining the respective strengths of each data source has not been documented to the best of our knowledge. We extend a methodology recently introduced in the flu surveillance literature, to combine dengue-related Google searches with historical dengue case counts to track dengue activity. Specifically, we evaluate the predictive performance of the ARGO model, as introduced in \cite{yang2015accurate}, to track dengue in five countries/regions around the globe: Mexico, Brazil, Thailand, Singapore, and Taiwan. Our contribution shows that the lessons learned to track influenza in data-rich environments, like the United States, can be used to develop methodologies to track an often-neglected tropical disease, dengue, in data-poor environments.

\section*{Materials and Methods}
\subsection*{Data}
We used two kinds of data sets for our study: (a) historical dengue incidence from government-led health agencies and (b) Google search fractions of dengue-related queries, aggregated at the national-level.

\subsubsection*{Historical dengue data}

\paragraph{Mexico.} 
Monthly-aggregated dengue case counts data from January 2001 to August 2015 were obtained from Mexico's Department of Epidemiology. \url{http://www.epidemiologia.salud.gob.mx/anuario/html/anuarios.html}

\paragraph{Brazil.}
Monthly dengue case counts data from January 2001 to December 2012 were obtained from the old website of Brazil's Ministry of Health (\url{http://dtr2004.saude.gov.br/sinanweb/tabnet/dh?sinannet/dengue/bases/denguebrnet.def}) on July 14, 2015. This site is no longer accessible, since the Ministry has moved to a new website (\url{http://portalsaude.saude.gov.br/index.php/situacao-epidemiologica-dados-dengue}), which now publishes weekly dengue data from 2014-present. This site contains annual dengue cases since 1990 but no longer has the historical monthly data. We confirmed that the annual totals match the sum of case counts over each year in our dataset. 

\paragraph{Thailand.}
Monthly dengue case count data from January 2003 to August 2015 were obtained from the Bureau of Epidemiology, Thailand (\url{http://www.boe.moph.go.th/boedb/surdata/disease.php?ds=66}). New data are published in an annual document available on the site.

\paragraph{Singapore.}
Weekly dengue case counts from January 10, 2004 to August 29, 2015 were obtained from the Singapore Ministry of Health and were aggregated into months. \url{https://www.moh.gov.sg/content/moh_web/home/statistics/infectiousDiseasesStatistics/weekly_infectiousdiseasesbulletin.html}. 

\paragraph{Taiwan.}
Weekly dengue case counts from January 3, 2009 to March 19, 2016 were obtained from the Taiwan Ministry of Health and Welfare and were aggregated into months. \url{http://nidss.cdc.gov.tw/ch/SingleDisease.aspx?dc=1&dt=4&disease=061&position=1}. 

\subsubsection*{Online search volume data}
Google search fractions for dengue-related queries were obtained from Google Trends (\url{www.google.com/trends}). 

\paragraph{Online search term selection}
We initially intended to use Google Correlate (\url{www.google.com/correlate}), which is designed to identify search terms correlating highly with a given time series, but we found the tool unreliable as many of the search terms returned were not related at all to dengue. Consequently, we used the Google Trends tool to identify the top ten queries most highly correlated with the term ‘dengue’ in each country, ignoring terms unrelated to dengue. The monthly aggregated search fractions were then downloaded within the time period of interest for each country. The query terms used for each country in this study are presented in Table S1, in the supporting material.

\subsection*{Methods}
We used the multivariate linear regression modeling framework ARGO (AutoRegressive model with GOogle search queries as exogenous variables)\cite{yang2015accurate}, previously used to track flu incidence using flu-related Google searches, to combine information from historical dengue case counts and dengue-related Google search frequencies with the goal of estimating dengue activity one month ahead of the publication of official local health reports. ARGO uses a training set that consists of a two-year moving time window (immediately prior to the month of prediction) and an $L_1$ regularization approach, to identify the best performing parsimonious model \cite{tibshirani1996regression}. This moving window approach allows the model to constantly improve its predictive ability by capturing the changing relationship between internet search behavior and dengue activity.

\subsubsection*{ARGO model formulation}
Let $y_t=\log(c_t + 1)$ be the \textit{log}-transformed dengue case counts $c_t$ at time $t$, and $X_{k,t}$ the \textit{log}-transformed Google search frequency of query term $k$ at time $t$.

Our ARGO model, derived from a hidden Markov model, is given by
\begin{equation}\label{eq:argo}
y_t =  \mu_y + \sum_{j \in J} \alpha_j y_{t-j} + \sum_{k \in K} \beta_k X_{k,t} + \epsilon_t,\quad \epsilon_t \overset{iid}{\sim} \mathcal{N}(0, \sigma^2),
\end{equation}
where $J$ is the set of auto-regressive lags, $K$ is the set of Google query terms, and $\bm X_t$ can be thought of as the exogenous variables to time series $\{y_t\}$ as introduced in \cite{yang2015accurate}.


\subsubsection*{ARGO model parameter estimation}
We take $J = \{1,\ldots,12\}\cup\{24\}$, i.e., $J$ consists of the most recent 12 months and the month exactly two years ago. Such choice of $J$ captures the influence of short and mid-term yearly fluctuations, as well as long-term seasonality previously shown to have strong predictive power in dengue \cite{johansson2016evaluating}. We take $K=10$, corresponding to the top ten dengue-related search terms as described in the data subsection.

We impose $L_1$ regularity for parameter estimation. In a given month, the goal is to find parameters $\mu_y$, $\bm \alpha= \{\alpha_j : j \in J\}$, and $\bm \beta= (\beta_1, ...,\beta_{10})$ that minimize
\begin{equation}\label{eq:ar_exogen_est}
\sum_t\left(y_t - \mu_y - \sum_{j \in J} \alpha_j y_{t-j} - \sum_{k=1}^{10} \beta_k X_{k,t} \right)^2 + \sum_{j\in J}\lambda_{\alpha_j}  |\alpha_j| + \sum_{k=1}^{10}\lambda_{\beta_k}  | \beta_k|
\end{equation}
where $\lambda_{\alpha_j},\lambda_{\beta_k}$ are regularization hyper-parameters.

For a given time window, ARGO automatically selects the most relevant variables to generate an out-of-sample dengue activity estimate. This is achieved by zeroing out regression coefficients of terms that contribute little (or have redundant information) to the prediction. This approach leads to interpretable results by allowing us to clearly identify which variables had a role in the prediction for each month or week of prediction.

All statistical analyses were performed with R, version 3.2.4. 

\subsection*{Benchmark models}
For comparison with ARGO, we included prediction results from five alternative methods. These are: 
\begin{enumerate} 
\item Google Dengue Trends \cite{chan2011using}, which ended in August 2015. Data are obtained from \url{https://www.google.org/flutrends/about/}. Because Google Dengue Trends reported dengue intensity in a scale from 0 to 1, we dynamically rescaled it using a sliding training window to recreate case estimates.
\item Penalized multivariate linear regression model with Google Trends information only \cite{santillana2014can}, denoted as GT. This is essentially ARGO without autoregressive lags, and incorporates a common $L_1$ penalty on the Google Trends data; 

\item Seasonal autoregressive model, denoted as SAR, using a time series of the most recent 3 lags, as well as 2 seasonal lags. Specifically, the monthly time series model is comprised of time lags 1,2,3,12,24: $y_t = \alpha_1 y_{t-1} + \alpha_2 y_{t-2} + \alpha_3 y_{t-3} + \alpha_{12} y_{t-12} + \alpha_{24} y_{t-24} + \epsilon_t$, $\epsilon_t\sim\mathcal{N}(0,\sigma^2)$.

\item Seasonal autoregressive model with Google Dengue Trends as exogenous variable, denoted as SAR+GDT. $y_t = \alpha_1 y_{t-1} + \alpha_2 y_{t-2} + \alpha_3 y_{t-3} + \alpha_{12} y_{t-12} + \alpha_{24} y_{t-24} + \beta\log\mathrm{GDT}_t + \epsilon_t$, $\epsilon_t\sim\mathcal{N}(0,\sigma^2)$.

\item Naive method, which simply uses the case count at the previous month as the prediction for the value of the current month.
\end{enumerate}
All benchmark models (except the naive method) were trained by linear regression with sliding two year windows for fair comparison.
\subsection*{Accuracy metrics}
We used five accuracy metrics to compare model performance: root mean squared error (RMSE), mean absolute error (MAE), root mean squared percentage error (RMSPE), mean absolute percentage error (MAPE), and Pearson correlation. 

Mathematically, these accuracy metrics of estimator $\hat c$ for target dengue case count $c$ are defined as, 
$\mathrm{RMSE}=\left[1/n\sum_{t=1}^n (\hat c_t - c_t)^2\right]^{1/2}$, 
$\mathrm{MAE}=1/n\sum_{t=1}^n |\hat c_t - c_t|$, 
$\mathrm{RMSPE}=\left\{1/n\sum_{t=1}^n [(\hat c_t - c_t)/c_t]^2\right\}^{1/2}$,
$\mathrm{MAPE}=1/n\sum_{t=1}^n {|\hat c_t - c_t|}/{c_t}$.
\subsection*{Retrospective predictions}
Retrospective out-of-sample estimates of dengue case counts were generated for each country using ARGO and the five benchmark models, assuming we only had access to information available at the time of prediction. 
The time windows in which we assessed the performance of our dengue estimates for each country were chosen based on the availability of official and GDT benchmark data.

These time windows are: Brazil from Mar 2006 -- Dec 2012, Mexico from Mar 2006 -- Aug 2015, Thailand from Oct 2010 -- Aug 2015, Singapore from Feb 2008 -- Aug 2015, and Taiwan from Jan 2013 -- Mar 2016.
\section*{Results}

\begin{figure}[!htbp]
\caption{{\bf Prediction results.}
Monthly dengue case-count predictions are displayed for all studied countries for four different prediction methodologies: ARGO, a seasonal auto-regressive model with and without Google Dengue Trends information (SAR+GDT, and SAR, respectively), and a naive prediction that estimates current month case counts using last month's observed cases. Historical dengue case counts, as reported by local health authorities, are shown for reference (black line), as well as the corresponding prediction errors associated to each methodology when compared to the reference.}
\label{fig:prediction}
\includegraphics[scale=0.75]{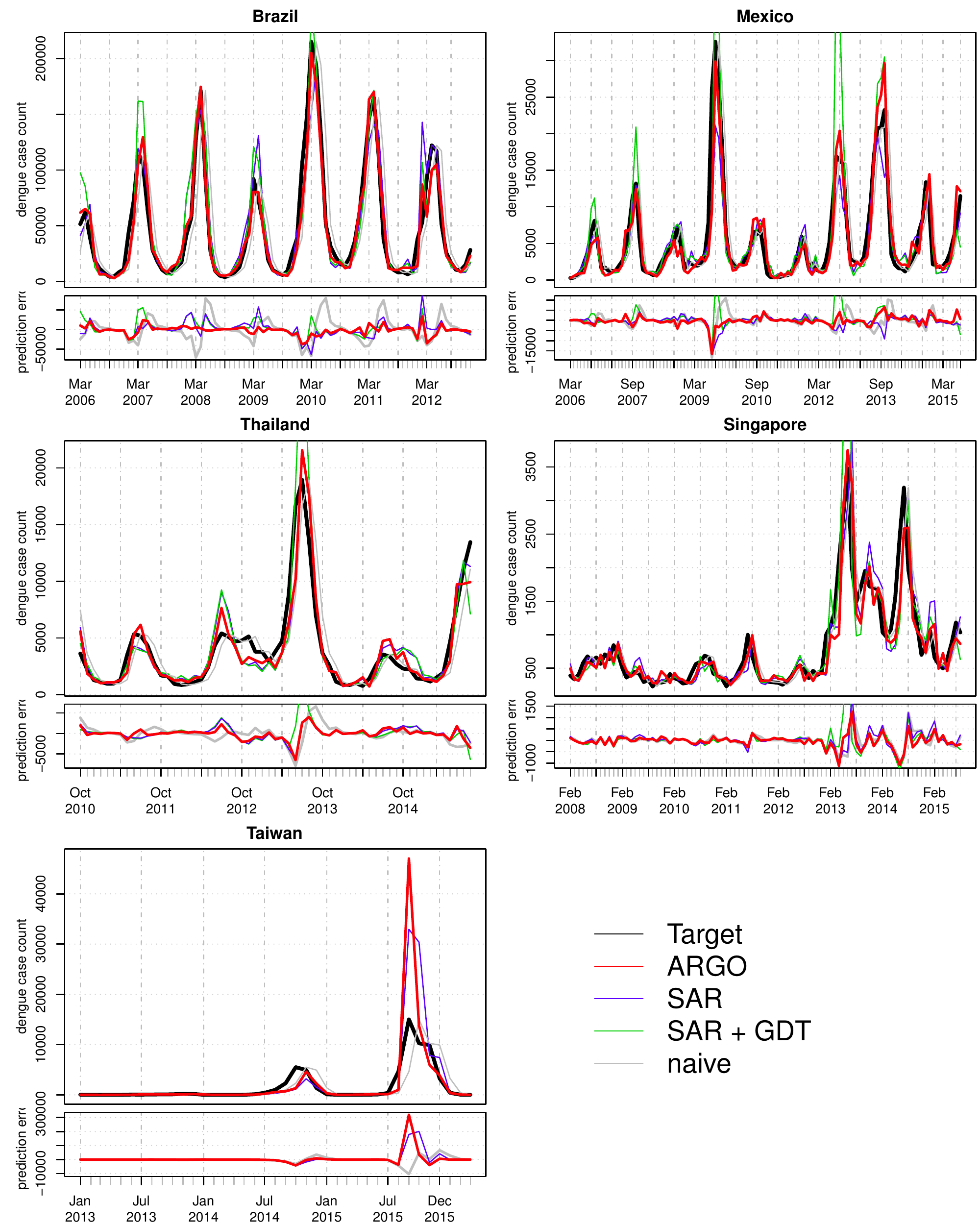}
\end{figure}

In four of the five countries/regions, Brazil, Mexico, Thailand and Singapore, ARGO outperformed all benchmark models across essentially all accuracy metrics (RMSE, MAE, RMSPE, MAPE, correlation). See Tab. \ref{tab:err}. In particular, by incorporating information from the internet searches and the historical dengue cases, ARGO achieved better results than using either information alone. This accuracy improvement is reflected in the accurate predictions during peaks of dengue activity and bounded estimates during off-season/periods with low levels of infection. See Fig. \ref{fig:prediction}. Unlike the seasonal autoregression with GDT model (SAR+GDT), ARGO avoided the significant overshooting problem that has been previously noted in Google Trends data (\cite{yang2015accurate}, \cite{santillana2014using}). This is especially notable between 2006-2008 and 2012-2014 in Mexico, and 2006-2010 in Brazil.

Taiwan shows notably different results. Of all the available models, the naive and seasonal autoregression models have the best performance, but neither is clearly effective. The naive model has the lowest RMSE and MAE, but the worst RMSPE, MAPE and correlation, while the seasonal model shows the best RMSPE, MAPE and correlation. In comparison, the other models have a much greater RMSE to MAE ratio, indicating worse performance during high prevalence relative to the naive model. ARGO does not outperform the benchmarks in this case.

This result seems to reflect the distinct case count pattern in Taiwan compared to the other countries. Taiwan experienced little to no dengue prevalence for years until two epidemic spikes occurred in 2014 and 2015. In contrast, the other countries experience seasonal fluctuations of dengue over their entire prediction windows. This lack of predictability may be the reason that both seasonal and Google information-based models have greater error than the naive model, significantly over-predicting the 2015 peak for example. Yet overall, these methods show greater correlation than the naive method, perhaps because they are more responsive. Because ARGO over-predicts to a greater extent than the autoregressive methodology, this again reflects previous observations on the tendency of Google data to overshoot.

\begin{table}[ht!]
\centering
\caption{{\bf Comparison of ARGO to benchmark models across countries and evaluation metrics} The bold face value is the best value among all methods according to each performance metric. Google Dengue Trends was not published for Taiwan and therefore the GDT benchmark is not available for Taiwan. The assessment period for the five regions, chosen based on the common available periods for all methods, are: Brazil (Mar 2006 -- Dec 2012), Mexico (Mar 2006 -- Aug 2015), Thailand (Oct 2010 -- Aug 2015), Singapore (Feb 2008 -- Aug 2015), Taiwan (Jan 2013 -- Mar 2016). The error value is relative to the naive, whose absolute error value is reported in the parenthesis.}\label{tab:err}

\begin{tabular}{rlllll}
  \hline
 & RMSE & MAE & RMSPE & MAPE & CORR \\ 
  \hline
\textbf{Brazil}\hfill\vadjust{} &  &  &  &  &  \\ 
  ARGO & \textbf{0.394} & \textbf{0.369} & \textbf{0.397} & \textbf{0.389} & \textbf{0.971} \\ 
  GDT & 0.666 & 0.633 & 0.984 & 0.817 & 0.916 \\ 
  GT & 0.902 & 0.829 & 0.877 & 0.838 & 0.861 \\ 
  SAR & 0.660 & 0.563 & 0.664 & 0.583 & 0.917 \\ 
  SAR+GDT & 0.629 & 0.587 & 0.564 & 0.560 & 0.938 \\ 
  naive & 1 (30560.436) & 1 (21677.634) & 1 (0.703) & 1 (0.546) & 0.812 \\ 
   \textbf{Mexico}\hfill\vadjust{} &  &  &  &  &  \\ 
   ARGO & \textbf{0.680} & \textbf{0.651} & \textbf{0.558} & \textbf{0.678} & \textbf{0.924} \\ 
   GDT & 0.944 & 0.961 & 1.270 & 1.311 & 0.863 \\ 
   GT & 0.950 & 0.927 & 1.097 & 1.100 & 0.861 \\ 
   SAR & 0.790 & 0.737 & 0.776 & 0.815 & 0.911 \\ 
   SAR+GDT & 1.249 & 0.986 & 0.779 & 0.854 & 0.891 \\ 
   naive & 1 (3570.105) & 1 (2161.018) & 1 (0.816) & 1 (0.492) & 0.833 \\ 
    \textbf{Thailand}\hfill\vadjust{} &  &  &  &  &  \\ 
    ARGO & \textbf{0.715} & \textbf{0.715} & \textbf{0.708} & \textbf{0.706} & \textbf{0.928} \\ 
    GDT & 0.880 & 0.868 & 1.494 & 1.284 & 0.884 \\ 
    GT & 1.364 & 1.224 & 1.510 & 1.368 & 0.833 \\ 
    SAR & 0.774 & 0.836 & 0.906 & 0.898 & 0.917 \\ 
    SAR+GDT & 1.157 & 0.983 & 0.923 & 0.936 & 0.903 \\ 
    naive & 1 (2058.891) & 1 (1276.068) & 1 (0.426) & 1 (0.326) & 0.852 \\ 
     \textbf{Singapore}\hfill\vadjust{} &  &  &  &  &  \\ 
     ARGO & \textbf{0.893} & \textbf{0.889} & \textbf{0.931} & \textbf{0.917} & \textbf{0.903} \\ 
     GDT & 1.182 & 1.285 & 1.427 & 1.439 & 0.821 \\ 
     GT & 1.287 & 1.165 & 1.287 & 1.254 & 0.796 \\ 
     SAR & 1.153 & 1.104 & 1.166 & 1.087 & 0.847 \\ 
     SAR+GDT & 2.452 & 1.297 & 1.185 & 1.009 & 0.775 \\ 
     naive & 1 (329.318) & 1 (202.651) & 1 (0.283) & 1 (0.230) & 0.878 \\ 
      \textbf{Taiwan}\hfill\vadjust{} &  &  &  &  &  \\ 
      ARGO & 2.180 & 1.264 & \textbf{0.233} & \textbf{0.359} & 0.834 \\ 
      GT & 12.211 & 4.904 & 1.069 & 0.898 & 0.724 \\ 
      SAR & 1.852 & 1.397 & 0.247 & 0.408 & \textbf{0.878} \\ 
      naive & \textbf{1 (2422.559)} & \textbf{1 (1063.597)} & 1 (3.248) & 1 (1.601) & 0.734 \\ 
   \hline
\end{tabular}
\end{table}

\section*{Discussion and conclusion}

Our findings confirm that combining historical dengue incidence information with dengue-related Google search data, in a self-adjusting manner, leads to better near real-time dengue activity estimates than those obtained with previous methodologies that exploit the information separately. 

ARGO's uniform out-performance of other benchmark methods for Mexico, Brazil, Thailand, and Singapore demonstrates its robustness and broad applicability. ARGO achieves this by balancing the influence of Internet search data, which quickly change in the face of outbreaks, and auto-regressive information, which tempers the predictions to mitigate the problem of overshooting. The application of an $L_1$ penalty helps identify the query terms most relevant to prediction at any given time, providing easy-to-interpret information. ARGO dynamically trains on a two-year rolling window, allowing model parameters to adjust over time to account for changes in Internet users' behavior. 

Our methodology works best in Mexico, Brazil, and Thailand, where a clear seasonal pattern is observed in the disease incidence trends over time. While the seasonality in Singapore is not as clear, ARGO still shows better performance than the alternatives. On the other hand, the results from Taiwan illustrate the limitations of our approach. The time series information does not contain an observable seasonal trend, and the model that uses only Google search terms as input does not reliably track dengue. This may be due to low case counts in Taiwan resulting in low dengue-related Internet search activity. In addition, the unpredictable nature of the outbreaks observed in Taiwan in the past couple of years may contribute to the poor performance of ARGO, and generally of all the methods considered in our comparison.

The quality of Google search data, and thus its ability to predict dengue activity, varies among countries. ARGO shows the strongest improvement over the seasonal autoregressive model in Brazil and Singapore, followed by Mexico and then Thailand. These differences in the effectiveness of Google query data across countries might be explained by the prevalence of the dengue virus, as observed in \cite{gluskin2014evaluation}, or by Internet penetration. The improvement of ARGO over its SAR counterpart is modest in Thailand, which could be potentially explained by lower internet accessibility.

We can intuitively expect that the more people search for dengue-related information during an outbreak, the more likely Google query information will be useful at predicting dengue activity. Although Taiwan has reasonably high internet penetration, the dengue case count is so low that dengue-related searches motivated by other medical or educational purposes may be introducing significant noise in the Google-query data.

The ultimate goal of this effort is to eventually build an accurate, real-time predictive modeling platform or disease barometer, where dengue case estimates are constantly updated to provide authorities and non-governmental organizations with valuable, actionable, close to real-time data on which they can make informed decisions, as well as providing travelers with vital information if they are visiting a high-risk area. This web- and app-based barometer would bring together multiple information sources in one place, including but not limited to traditional epidemiological case reports, Google search data, social media data, crowd-sourced data, climate and transportation data etc., and integrates a rapid response and alert system for users based on their specific location. 

The platform would also enable users to verify dengue risk information with their own observations, creating a positive feedback loop that will continuously improve the accuracy of the tool and its predictions. We are currently building such a platform at \url{Healthmap.org/denguetrends}, and through Break Dengue’s ``Dengue Track'' initiative \url{www.breakdengue.org/dengue-track/}, a crowd-sourced tool currently in beta that offers a user-friendly online chat system which maps dengue cases worldwide, and gives the public free access to tool-kits that help reduce their risk of infection. The potential impact would be far reaching, as the same model could also be used to track and map countless other infectious and mosquito-borne diseases, like zika, malaria, yellow fever or Chikungunya.

In light of ARGO's strengths and limitations, further efforts on dengue activity estimation may focus on, for example, producing short-term forecasts of dengue activity, in addition to the nowcast presented here (See \cite{santillana2015combining} for such an extension for flu forecasting). Our approach may help produce dengue activity estimates in higher spatial resolutions that can lead to alert systems for people with an increased risk of exposure to the dengue virus at any given point in time. It is important to keep in mind that state-level or city-level spatial scales with low dengue activity may present similar challenges to the applicability of our approach as seen in Taiwan. The incorporation of other Internet-based data sources \cite{santillana2015combining} and cross-country spatial relationships should also be exploited in order to improve the accuracy in predictions.


\newpage
\section*{Supporting Information}

\begin{table}[ht!]
\centering
\caption{\textbf{Query terms used for each country and region}}
\label{tab:query}
\centering
\includegraphics[scale=0.8]{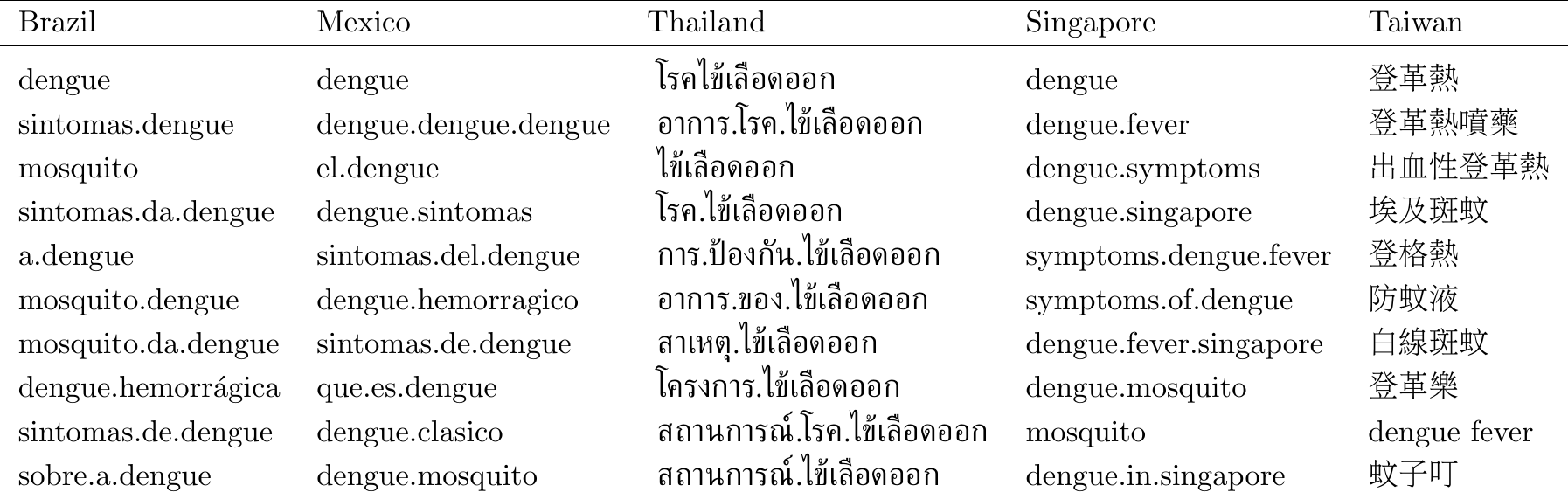}

\end{table}

\subsection*{ARGO hyper-parameters for each country and region}
\paragraph*{Mexico}
Since we found the nearest three time lags to have significant predictive effect on future dengue occurrence, we decided not to penalize these three time lags, setting $\lambda_{\alpha_j}=0$, $j = 1,2,3$. We do not have knowledge of the predictive power of the later time lags, so we set a common penalty for all of them $\lambda_{\alpha_j}=\lambda_\alpha$, $j \ge 4$. We applied the same argument to the Google search terms. We further set $\lambda_\alpha= \lambda_\beta$ to reduce the number of hyper parameters, and $\lambda_{\alpha_k}=\lambda_{\beta_k}=0$, $k = 1,2,3$, $\lambda_{\alpha_k}=\lambda_{\beta_k}=\lambda$, $k = 4,\ldots, 10$. 
\paragraph*{Brazil}
We found the same pattern for Brazil as for Mexico. Thus, we set $\lambda_{\alpha_k}=\lambda_{\beta_k}=0$, $k = 1,2,3$, $\lambda_{\alpha_k}=\lambda_{\beta_k}=\lambda$, $k = 4,\ldots, 10$.
\paragraph*{Thailand}
The first three time lags for Thailand were significant, but none of the Google terms by itself are significant. This observation inspires us to set the hyper-parameters as $\lambda_{\alpha_j}=0$, $j = 1,2,3$, $\lambda_{\alpha_j}=\lambda$, $j = 4,\ldots,10$, $\lambda_{\beta_k}=\lambda$, $k=1,\ldots, 10$.
\paragraph*{Singapore}
Singapore has similar pattern with Thailand, so we set $\lambda_{\alpha_j}=0$, $j = 1,2,3$, $\lambda_{\alpha_j}=\lambda$, $j = 4,\ldots,10$, $\lambda_{\beta_k}=\lambda$, $k=1,\ldots, 10$.
\paragraph*{Taiwan}
We found that Taiwan data and the Google search terms for Taiwan had similar pattern to that of Singapore, so we set $\lambda_{\alpha_j}=0$, $j = 1,2,3$, $\lambda_{\alpha_j}=\lambda$, $j = 4,\ldots,10$, $\lambda_{\beta_k}=\lambda$, $k=1,\ldots, 10$.

\subsection*{Aggregate from weekly data to monthly data}
We aggregate the Google Trends data from weekly frequency to monthly frequency using summation. If a fraction of week belongs to certain month, the summing value will be the fraction times the value reported for that week.

\subsection*{Robustness to Google Trends variation}
We include a robustness study to identify the effects of the observed variations in the (input) data acquired from the Google Trends website. For this, we downloaded 11 copies of data on different days in November 2016, and repeated the implementation of the methodology described in the main text. Our findings are presented in Table \ref{tab:gt_variation}. The mean of the 11 evaluation metric values is displayed as well as the standard deviation, in parenthesis. GDT has no variation since it is taken as exogenous in this study. If we had access to the raw data GDT is constructed from, we should expect to see similar variations as well. Autoregressive models do not suffer from these variation since they do not use Google Trends data as input. As expected, ARGO, which combines Google Trends data with time series data, suffers less from the variations of the Google Trends data than the model based on Google Trends data only.

\begin{table}[ht]
\centering

\caption{\textbf{Sensitivity to Google Trends variation.} The mean evaluation metric value of the 11 different datasets is displayed in the table, as well as the the standard deviation. All values are absolute.}
\label{tab:gt_variation}
\small
\begin{tabular}{rlllll}
  \hline
 & RMSE & MAE & RMSPE & MAPE & CORR \\ 
  \hline
\textbf{Brazil}\hfill\vadjust{} &  &  &  &  &  \\ 
  ARGO & 14602.591(1303.123) & 9043.447(746.341) & 0.329(0.029) & 0.234(0.014) & 0.957(0.008) \\ 
  GDT & 20349.593(0) & 13725.535(0) & 0.692(0) & 0.446(0) & 0.916(0) \\ 
  GT & 31606.088(3458.28) & 20243.862(1716.244) & 0.821(0.07) & 0.565(0.043) & 0.815(0.04) \\ 
  SAR & 20158.471(0) & 12215.217(0) & 0.467(0) & 0.318(0) & 0.917(0) \\ 
  SAR+GDT & 19220.295(0) & 12732.517(0) & 0.397(0) & 0.306(0) & 0.938(0) \\ 
  naive & 30560.436(0) & 21677.634(0) & 0.703(0) & 0.546(0) & 0.812(0) \\ 
   \textbf{Mexico}\hfill\vadjust{} &  &  &  &  &  \\ 
   ARGO & 2695.046(145.838) & 1532.008(79.432) & 0.516(0.063) & 0.355(0.025) & 0.903(0.011) \\ 
   GDT & 3370.184(0) & 2076.24(0) & 1.036(0) & 0.645(0) & 0.863(0) \\ 
   GT & 4628.805(456.821) & 2528.918(208.179) & 1.016(0.13) & 0.616(0.045) & 0.705(0.065) \\ 
   SAR & 2821.504(0) & 1593.552(0) & 0.633(0) & 0.401(0) & 0.911(0) \\ 
   SAR+GDT & 4460.343(0) & 2131.342(0) & 0.635(0) & 0.42(0) & 0.891(0) \\ 
   naive & 3570.105(0) & 2161.018(0) & 0.816(0) & 0.492(0) & 0.833(0) \\ 
    \textbf{Thailand}\hfill\vadjust{} &  &  &  &  &  \\ 
    ARGO & 1543.473(129.498) & 911.561(43.288) & 0.303(0.014) & 0.23(0.008) & 0.925(0.011) \\ 
    GDT & 1811.26(0) & 1107.728(0) & 0.636(0) & 0.419(0) & 0.884(0) \\ 
    GT & 2590.984(499.302) & 1582.48(134.678) & 0.687(0.068) & 0.495(0.04) & 0.82(0.05) \\ 
    SAR & 1592.675(0) & 1066.51(0) & 0.386(0) & 0.293(0) & 0.917(0) \\ 
    SAR+GDT & 2381.833(0) & 1253.851(0) & 0.393(0) & 0.305(0) & 0.903(0) \\ 
    naive & 2058.891(0) & 1276.068(0) & 0.426(0) & 0.326(0) & 0.852(0) \\ 
     \textbf{Singapore}\hfill\vadjust{} &  &  &  &  &  \\ 
     ARGO & 309.492(24.395) & 185.639(7.578) & 0.282(0.011) & 0.22(0.005) & 0.895(0.014) \\ 
     GDT & 389.389(0) & 260.421(0) & 0.404(0) & 0.331(0) & 0.821(0) \\ 
     GT & 362.286(30.443) & 246.596(13.725) & 0.398(0.019) & 0.323(0.017) & 0.866(0.031) \\ 
     SAR & 379.794(0) & 223.633(0) & 0.33(0) & 0.25(0) & 0.847(0) \\ 
     SAR+GDT & 807.414(0) & 262.783(0) & 0.336(0) & 0.232(0) & 0.775(0) \\ 
     naive & 329.318(0) & 202.651(0) & 0.283(0) & 0.23(0) & 0.878(0) \\ 
      \textbf{Taiwan}\hfill\vadjust{} &  &  &  &  &  \\ 
      ARGO & 2919.016(1284.247) & 989.77(258.632) & 0.846(0.154) & 0.628(0.062) & 0.873(0.026) \\ 
      GT & 5031.846(7248.156) & 1336.656(1157.202) & 4.092(1.126) & 1.655(0.272) & 0.848(0.062) \\ 
      SAR & 4487.372(0) & 1485.911(0) & 0.801(0) & 0.653(0) & 0.878(0) \\ 
      naive & 2422.559(0) & 1063.597(0) & 3.248(0) & 1.601(0) & 0.734(0) \\ 
   \hline
\end{tabular}
\end{table}

\clearpage
\section*{Acknowledgments}

\end{document}